\RequirePackage{fix-cm}
\documentclass[twocolumn,epjc3,final]{svjour3}  
\smartqed  
 \usepackage{mathrsfs}
\usepackage{amsmath, txfonts}
\usepackage{graphicx,bm,booktabs,bm}
\usepackage{longtable,lscape}
\usepackage{overpic,multirow,color}
\definecolor{cover}{rgb}{0.77,0.87,0.88}
\definecolor{blueone}{rgb}{0.1,0.1,.7}
\definecolor{citec}{rgb}{0.14,0.47,0.09}
\definecolor{two}{rgb}{0.0,0.5,0.}
\definecolor{three}{rgb}{.5,.1,0.15}
\usepackage[bookmarks=true,bookmarksopen=false,plainpages=false,breaklinks=true,
   bookmarksnumbered=true,hypertexnames=false,
   filecolor=blue,urlcolor=three,menucolor=three,
   linkcolor=three,citecolor=blueone, colorlinks,
   anchorcolor=blue,runcolor=pink,frenchlinks=red
   pdfstartview=FitH,pdftitle=title,%
   pdfauthor=author]{hyperref}
              \allowdisplaybreaks[4]

\graphicspath{{Figures/}} %

\journalname{Eur. Phys. J. C}
\begin{document}
\title{Molecular states from $\Sigma^{(*)}_c\bar{D}^{(*)}-\Lambda_c\bar{D}^{(*)}$  interaction}
\author{Jun He\thanksref{e1,addr1}
\and Dian-Yong Chen\thanksref{addr2}
}                     
\thankstext{e1}{Corresponding author: junhe@njnu.edu.cn}
\institute{Department of  Physics and Institute of Theoretical Physics, Nanjing Normal University,
Nanjing 210097, China\label{addr1}
\and
School of Physics, Southeast University, Nanjing 210094,  China\label{addr2}
}

\date{Received: date / Revised version: date}
%
\maketitle

\abstract{
In this work, we systemically investigate the molecular states from the $\Sigma^{(*)}_c\bar{D}^{(*)}-\Lambda_c\bar{D}^{(*)}$  interaction with the help of the Lagrangians with heavy quark and chiral symmetries in a quasipotential Bethe-Salpeter equation (qBSE) approach.  The molecular  states   are produced from isodoublet (I=1/2) $\Sigma_c\bar{D}$ interaction with spin parity $J^P=1/2^-$ and $\Sigma_c\bar{D}^*$ interaction with  $1/2^-$ and $ 3/2^-$. Their masses and widths  are  consistent with the $P_c(4312)$, $P_c(4440)$ and $P_c(4457)$  observed at LHCb.  The  states, $\Sigma_c^*\bar{D}^*(1/2^-)$,  $\Sigma_c^*\bar{D}^*(3/2^-)$ and  $\Sigma^*_c\bar{D}(3/2^-)$, are also produced with the same parameters.  The isodoublet $\Sigma_c^*\bar{D}^*$ interaction with $5/2^-$, as well as  the isoquartet (I=3/2) $\Sigma_c\bar{D}^*$ interactions  with $1/2^-$ and $3/2^-$, $\Sigma_c^*\bar{D}^*$ interaction with $3/2^-$ and $5/2^-$,  are also attractive while very large cutoff is required to produce a molecular state. We also investigate the origin of the widths of these molecular states in the same qBSE frame.  The $\Lambda\bar{D}^*$ channel is dominant in the decays of the states, $\Sigma_c\bar{D}^*(1/2^-)$, $\Sigma_c\bar{D}^*(3/2^-)$, $\Sigma_c^*\bar{D}(3/2^-)$, and $\Sigma_c\bar{D}(1/2^-)$.  The $\Sigma^*_c\bar{D}^*(1/2^-)$ state has large coupling to $\Sigma_c\bar{D}$ channel while  the $\Sigma_c\bar{D}^*$, $\Sigma^*_c\bar{D}$ and $\Lambda_c\bar{D}^*$ channels provide similar contributions to the width of the $\Sigma^*_c\bar{D}^*(3/2^-)$ state. These results will be helpful to understand the current LHCb experimental results,  and the three predicted states and the decay pattern of these hidden-charmed molecular pentaquarks can be checked in  future experiments. 
} 

\section{Introduction}

The  study of  exotic hadrons is an important topic in understanding how quarks combine to a hadron.  One  type of the exotic hadrons is the molecular state, which is a shallow bound state of two and more hadrons. Though it is not so fancy as a compact multiquark,  it seems easier to be produced because its constituent hadrons are realistic.  In the side of experiment, many XYZ particles were observed near the threshold of charmed-anticharmed or bottom-antibottom mesons. For example, the $X(3872)/Z_c(3900)$, $Z_c(4020)$,  $Z_b(10610)$ and $Z_b(10650)$ are very close to the $D\bar{D}^*$, $D^*\bar{D}^*$, $B\bar{B}^*$ and $B^*\bar{B}^*$ thresholds, respectively. It suggests that such particles are from the interactions of the corresponding hadrons. It also makes the molecular state picture become a popular interpretation of the XYZ particles.

Recently, the LHCb Collaboration updated their observation of  pentaquark candidates~\cite{Aaij:2019vzc}.  The upper one, $P_c(4450)$,  resolves into two resonances, $P_c(4440)$ and $P_c(4457)$, and a new pentaquark, $P_c(4312)$, was observed near the $\Sigma_c\bar{D}$ threshold. The $P_c(4380)$ reported in the previous observation~\cite{Aaij:2015tga} is suspended to wait construction of new amplitude model.  
The four pentaquarks, $P_c(4457)$, $P_c(4440)$, $P_c(4380)$, and $P_c(4312)$, construct a good pattern for all S-wave molecular states from $\Sigma_c\bar{D}^*$, $\Sigma^*_c\bar{D}$, and $\Sigma_c\bar{D}$ interactions, which has been predicted partly in the literature~\cite{Wu:2010jy,Wang:2011rga,Yang:2011wz,Wu:2012md}.  After the LHCb results released, many theoretical interpretations in the molecular state picture were proposed~\cite{Chen:2015loa,Chen:2015moa,Karliner:2015ina,Roca:2015dva,He:2015cea,Burns:2015dwa,He:2016pfa,Chen:2019asm,Chen:2019bip,Fernandez-Ramirez:2019koa,Wu:2019rog,Wang:2018waa,Huang:2019jlf,Huang:2015uda}.

To further confirm the molecular state interpretation of the $P_c$ states, it is very helpful to make a prediction of more states with the relevant interactions. The four $P_c$ states observed at LHCb are  all from the $\Sigma_c^{(*)}\bar{D}^{(*)}$ interaction. If we only consider the S-wave states, there should be seven possible molecular states. The states from the $\Sigma_c\bar{D}$, $\Sigma_c^*\bar{D}$, and $\Sigma_c\bar{D}^*$ interactions have been filled by the experimental observed $P_c(4312)$, $P_c(4380)$, $P_c(4440)$, and $P_c(4457)$.  It is interesting to find out if there exist  three S-wave $\Sigma^*\bar{D}^*$ molecular states.  In Refs.~\cite{Liu:2019tjn,Liu:2019zvb,Pan:2019skd}, such states has been studied in a parameterized model, and the authors suggested the existence of three  $\Sigma^*\bar{D}^*$ states with the observed $P_c$ states as input. 

Theoretically,  a state usually  exhibits different decay behaviors in different theoretical pictures.   The decay pattern of the pentaquarks is another way to check their internal structure.    In the literature, the decays of the $P_c$ states have been studied by many authors~\cite{Lin:2018kcc,Lin:2019qiv,Xiao:2019mst}.  In Ref.~\cite{Lin:2019qiv}, it was suggested that the $\Lambda_c\bar{D}^*$ channel is very important in the decays of the $P_c$ states.   The mass and the decay pattern of a molecular state is often studied separately in the literature. If the $P_c$ states are molecular states from the $\Sigma_c^{(*)}\bar{D}^{(*)}$ interaction, the decays to these channels can be obtained as a coproduct after the coupled-channel effect is included.  A bound state will acquire a width, and exhibits itself as a pole in the complex plane after adding another interaction channel below the production channel to make a coupled-channel calculation.  Hence, it is interesting to study the production of the molecular states produced from the interaction and their decay behaviors in the same theoretical frame.

In our previous work~\cite{He:2019ify}, we studied the $\Sigma^{(*)}_c\bar{D}^{(*)}$ interaction and focused on the molecular states which can be related to the three pentaquarks observed at LHCb. A calculation in a quasipotential Bethe-Salpeter equation (qBSE) approach suggests that the $P_c(4457)$, $P_c(4440)$ and $P_c(4312)$ can be explained as two $\Sigma_c \bar{D}^*$ molecular states with $3/2^-$ and $1/2^-$ and a $\Sigma_c \bar{D}$ molecular state with $1/2^-$.  An enhancement was also found near the  $\Sigma_c^*\bar{D}$  threshold with $3/2^-$.  It is naturally to extend the calculation to study all possible S-wave isodoublet ($I=1/2$)molecular states from the $\Sigma^{(*)}_c\bar{D}^{(*)}$ interaction  and their isoquartet ($I=3/2$) partners.  The previous calculation suggests that the coupled-channel effect between different channels is  small for the molecular states related to the $P_c(4312)$, $P_c(440)$, and $P_c(4457)$, which leads to very small widths if only $\Sigma^{(*)}_c\bar{D}^{(*)}$ interaction considered. The widths of those states are possible from the coupling to  $\Lambda_c\bar{D}^*$ channel as suggested in Ref.~\cite{Lin:2019qiv}.  Hence, in the current work, we will make a systemically calculation of the $\Sigma^{(*)}_c\bar{D}^{(*)}-\Lambda_c\bar{D}^{(*)}$  interaction in the qBSE approach to find out all possible S-wave molecular states and to study the couplings of these channels in the same frame. 

This work is organized as follows. After introduction, the detail of the dynamical study of coupled-channel $\Sigma^{(*)}_c\bar{D}^{(*)}-\Lambda_c\bar{D}^{(*)}$   interactions will be presented, which includes relevant effective Lagrangians, reduction of potential kernel and a brief introduction of the qBSE. Then, the results of shingle-channel calculation  are given in Section 3 to present the possible bound states produced form the $\Sigma^{(*)}_c\bar{D}^{(*)}-\Lambda_c\bar{D}^{(*)}$ interaction.  The coupled-channel results are presented in section 4.  The bound states obtained in Section 3 become poles in complex plane, which will be compared with the experimental results. The poles of the molecular states from full coupled-channel and  two-channel calculations will be presented also, which can be related to their decay widths.  Finally, summary  and  discussion will be given in the last section.

\section{Theoretical frame}

To study the bound states from the $\Sigma^{(*)}_c\bar{D}^{(*)}-\Lambda_c\bar{D}^{(*)}$  interaction and the couplings between different channels, we need to construct the coupled-channel potential kernel.  In the current work, we adopt the one-boson-exchange model to describe the interaction between the charmed baryon and anticharmed meson. The exchanges by peseudoscalar $\mathbb{P}$, vector $\mathbb{V}$ and $\sigma$ mesons will be considered.  Hence, the effective Lagrangians depicting the couplings of light mesons and  anti-charmed mesons or charmed baryons  are required and will be presented in the below.

\subsection{Relevant Lagrangians}
First, we consider the couplings of light mesons to heavy-light anticharmed mesons $\tilde{\mathcal{P}}=(\bar{D}^0, D^-, D^-_s)$.  In terms of heavy quark limit and chiral symmetry,  the Lagrangians have been  constructed  in the literature as \cite{Cheng:1992xi,Yan:1992gz,Wise:1992hn,Casalbuoni:1996pg},
\begin{eqnarray}
\mathcal{L}_{HH\mathbb{P}}&=& ig_{1}\langle \bar{H}_a^{\bar{Q}} \gamma_\mu
A_{ba}^\mu\gamma_5 {H}_b^{\bar{Q}}\rangle,\label{eq:lag}\nonumber\\
\mathcal{L}_{HH\mathbb{V}}&=& -i\beta\langle \bar{H}_a^{\bar{Q}}  v_\mu
(\mathcal{V}^\mu_{ab}-V^\mu_{ab}){H}_b^{\bar{Q}}\rangle
+i\lambda\langle \bar{H}_b^{\bar{Q}}
\sigma_{\mu\nu}F^{\mu\nu}(\rho)\bar{H}_a^{\bar{Q}}\rangle,
\nonumber\\
\mathcal{L}_{ HH\sigma}&=&g_s \langle \bar{H}_a^{\bar{Q}}\sigma
\bar{H}_a^{\bar{Q}}\rangle,\label{eq:lag2}
\end{eqnarray}
where the axial current is
$A^\mu=\frac{1}{2}(\xi^\dag\partial_\mu\xi-\xi \partial_\mu
\xi^\dag)=\frac{i}{f_\pi}\partial_\mu{\mathbb P}+\cdots$ with
$\xi=\exp(i\mathbb{P}/f_\pi)$ and $f_\pi=132$
MeV.
$\mathcal{V}_\mu=\frac{i}{2}[\xi^\dag(\partial_\mu\xi)
+(\partial^\mu\xi)\xi^\dag]=0$.
$V^\mu_{ba}=ig_\mathbb{V}\mathbb{V}^\mu_{ba}/\sqrt{2}$, and
$F_{\mu\nu}(\rho)=\partial_\mu V_\nu - \partial_\nu V_\mu +
[\rho_\mu,{\ } \rho_\nu]$. The $\mathbb
P$ and $\mathbb V$ are the pseudoscalar and vector matrices as
\begin{eqnarray}
    {\mathbb P}&=&\left(\begin{array}{ccc}
        \frac{1}{\sqrt{2}}\pi^0+\frac{\eta}{\sqrt{6}}&\pi^+&K^+\\
        \pi^-&-\frac{1}{\sqrt{2}}\pi^0+\frac{\eta}{\sqrt{6}}&K^0\\
        K^-&\bar{K}^0&-\frac{2\eta}{\sqrt{6}}
\end{array}\right),\nonumber\\
\mathbb{V}&=&\left(\begin{array}{ccc}
\frac{\rho^{0}}{\sqrt{2}}+\frac{\omega}{\sqrt{2}}&\rho^{+}&K^{*+}\\
\rho^{-}&-\frac{\rho^{0}}{\sqrt{2}}+\frac{\omega}{\sqrt{2}}&K^{*0}\\
K^{*-}&\bar{K}^{*0}&\phi
\end{array}\right).\label{MPV}
\end{eqnarray}
The $H_a^{\bar{Q}}=[\tilde{\mathcal{P}}^{*\mu}_a\gamma_\mu-\tilde{\mathcal{P}}_a\gamma_5]\frac{1-\rlap\slash
v}{2}$ and
$\bar{H}=\gamma_0H^\dag\gamma_0$ with $v=(1,\mathbf{0})$. The
$\tilde{\mathcal{P}}
$ and $\tilde{\mathcal{P}}^*
$ satisfy the normalization relations $\langle
0|\tilde{{\mathcal{P}}}|\bar{Q}{q}(0^-)\rangle
=\sqrt{M_\mathcal{P}}$ and $\langle
0|\tilde{{\mathcal{P}}}^*_\mu|\bar{Q}{q}(1^-)\rangle=
\epsilon_\mu\sqrt{M_{\mathcal{P}^*}}$. 

The Lagrangians can be further expanded
as follows,
\begin{eqnarray}\label{eq:lag-p-exch}
  \mathcal{L}_{\tilde{\mathcal{P}}^*\tilde{\mathcal{P}}\mathbb{P}} &=&
 i\frac{2g\sqrt{m_{\tilde{\mathcal{P}}} m_{\tilde{\mathcal{P}}^*}}}{f_\pi}
  (-\tilde{\mathcal{P}}^{*\dag}_{a\lambda}\tilde{\mathcal{P}}_b
  +\tilde{\mathcal{P}}^\dag_{a}\tilde{\mathcal{P}}^*_{b\lambda})
  \partial^\lambda\mathbb{P}_{ab},\nonumber\\
    \mathcal{L}_{\tilde{\mathcal{P}}^*\tilde{\mathcal{P}}^*\mathbb{P}} &=&
-\frac{g}{f_\pi} \epsilon_{\alpha\mu\nu\lambda}\tilde{\mathcal{P}}^{*\mu\dag}_a
\overleftrightarrow{\partial}^\alpha \tilde{\mathcal{P}}^{*\lambda}_{b}\partial^\nu\mathbb{P}_{ba},\nonumber\\
    \mathcal{L}_{\tilde{\mathcal{P}}^*\tilde{\mathcal{P}}\mathbb{V}} &=&
\sqrt{2}\lambda g_V\varepsilon_{\lambda\alpha\beta\mu}
  (-\tilde{\mathcal{P}}^{*\mu\dag}_a\overleftrightarrow{\partial}^\lambda
  \tilde{\mathcal{P}}_b  +\tilde{\mathcal{P}}^\dag_a\overleftrightarrow{\partial}^\lambda
  \tilde{\mathcal{P}}_b^{*\mu})(\partial^\alpha{}\mathbb{V}^\beta)_{ab},\nonumber\\
	\mathcal{L}_{\tilde{\mathcal{P}}\tilde{\mathcal{P}}\mathbb{V}} &=& -i\frac{\beta	g_V}{\sqrt{2}}\tilde{\mathcal{P}}_a^\dag
	\overleftrightarrow{\partial}_\mu \tilde{\mathcal{P}}_b\mathbb{V}^\mu_{ab}, \nonumber\\
  \mathcal{L}_{\tilde{\mathcal{P}}^*\tilde{\mathcal{P}}^*\mathbb{V}} &=& - i\frac{\beta
  g_V}{\sqrt{2}}\tilde{\mathcal{P}}_a^{*\dag}\overleftrightarrow{\partial}_\mu
  \tilde{\mathcal{P}}^*_b\mathbb{V}^\mu_{ab}\nonumber\\
  &-&i2\sqrt{2}\lambda  g_Vm_{\tilde{\mathcal{P}}^*}\tilde{\mathcal{P}}^{*\mu\dag}_a\tilde{\mathcal{P}}^{*\nu}_b(\partial_\mu\mathbb{V}_\nu-\partial_\nu\mathbb{V}_\mu)_{ab}
,\nonumber\\
  \mathcal{L}_{\tilde{\mathcal{P}}\tilde{\mathcal{P}}\sigma} &=&
  -2g_s m_{\tilde{\mathcal{P}}}\tilde{\mathcal{P}}_a^\dag \tilde{\mathcal{P}}_a\sigma, \nonumber\\
  \mathcal{L}_{\tilde{\mathcal{P}}^*\tilde{\mathcal{P}}^*\sigma} &=&
  2g_s m_{\tilde{\mathcal{P}^*}}\tilde{\mathcal{P}}_a^{*\dag}
  \tilde{\mathcal{P}}^*_a\sigma,\label{LD}
\end{eqnarray}
where  the $v$ is replaced by $i\overleftrightarrow{\partial}/\sqrt{m_im_f}$ with the $m_{i,f}$ is for the initial or final $\bar{D}^{(*)}$ meson.

The Lagrangians for the couplings between the charmed baryon and light mesons can also be constructed in the heavy quark limit and under chiral symmetry as,
\begin{eqnarray}
{\cal L}_{S}&=&-
\frac{3}{2}g_1(v_\kappa)\epsilon^{\mu\nu\lambda\kappa}{\rm tr}[\bar{S}_\mu
A_\nu S_\lambda]+i\beta_S{\rm tr}[\bar{S}_\mu v_\alpha (\mathcal{V}^\alpha-
V^\alpha)
S^\mu]\nonumber\\
&& + \lambda_S{\rm tr}[\bar{S}_\mu F^{\mu\nu}S_\nu]
+\ell_S{\rm tr}[\bar{S}_\mu \sigma S^\mu],\\
{\cal L}_{B_{\bar{3}}}&=& i\beta_B{\rm tr}[\bar{B}_{\bar{3}}v_\mu(\mathcal{V}^\mu-V^\mu)
B_{\bar{3}}]
+\ell_B{\rm tr}[\bar{B}_{\bar{3}}{\sigma} B_{\bar{3}}], \\
{\cal L}_{int}&=&ig_4 {\rm tr}[\bar{S}^\mu A_\mu B_{\bar{3}}]+i\lambda_I \epsilon^{\mu\nu\lambda\kappa}v_\mu{\rm tr}[\bar{S}_\nu F_{\lambda\kappa} B_{\bar{3}}]+h.c.,
\end{eqnarray}
where $S^{\mu}_{ab}$ is composed of Dirac spinor operators,
\begin{eqnarray}
    S^{ab}_{\mu}&=&-\sqrt{\frac{1}{3}}(\gamma_{\mu}+v_{\mu})
    \gamma^{5}B^{ab}+B^{*ab}_{\mu}\equiv{ B}^{ab}_{0\mu}+B^{ab}_{1\mu},\nonumber\\
    \bar{S}^{ab}_{\mu}&=&\sqrt{\frac{1}{3}}\bar{B}^{ab}
    \gamma^{5}(\gamma_{\mu}+v_{\mu})+\bar{B}^{*ab}_{\mu}\equiv{\bar{B}}^{ab}_{0\mu}+\bar{B}^{ab}_{1\mu},
\end{eqnarray}
and the  the charmed baryon matrices are defined as 
\begin{eqnarray}
B_{\bar{3}}&=&\left(\begin{array}{ccc}
0&\Lambda^+_c&\Xi_c^+\\
-\Lambda_c^+&0&\Xi_c^0\\
-\Xi^+_c &-\Xi_c^0&0
\end{array}\right),\quad
B=\left(\begin{array}{ccc}
\Sigma_c^{++}&\frac{1}{\sqrt{2}}\Sigma^+_c&\frac{1}{\sqrt{2}}\Xi'^+_c\\
\frac{1}{\sqrt{2}}\Sigma^+_c&\Sigma_c^0&\frac{1}{\sqrt{2}}\Xi'^0_c\\
\frac{1}{\sqrt{2}}\Xi'^+_c&\frac{1}{\sqrt{2}}\Xi'^0_c&\Omega^0_c
\end{array}\right).\label{MBB}
\end{eqnarray}

The explicit forms of the Lagrangians can be written as,
\begin{eqnarray}
{\cal L}_{BB\mathbb{P}}&=&i\frac{3g_1}{2f_\pi\sqrt{m_{\bar{B}}m_{B}}}~\epsilon^{\mu\nu\lambda\kappa}\partial^\nu \mathbb{P}~
\sum_{i=0,1}\bar{B}_{i\mu} \overleftrightarrow{\partial}_\kappa B_{j\lambda},\nonumber\\
{\cal L}_{BB\mathbb{V}}&=&-\frac{\beta_S g_V}{\sqrt{2m_{\bar{B}}m_{B}}}\mathbb{V}^\nu
 \sum_{i=0,1}\bar{B}_i^\mu \overleftrightarrow{\partial}_\nu B_{j\mu}\nonumber\\
&-&\frac{\lambda_S
g_V}{\sqrt{2}}(\partial_\mu \mathbb{V}_\nu-\partial_\nu \mathbb{V}_\mu) \sum_{i=0,1}\bar{B}_i^\mu B_j^\nu,\nonumber\\
{\cal L}_{BB\sigma}&=&\ell_S\sigma\sum_{i=0,1}\bar{B}_i^\mu B_{j\mu},\nonumber\\
    {\cal L}_{B_{\bar{3}}B_{\bar{3}}\mathbb{V}}&=&-\frac{g_V\beta_B}{\sqrt{2m_{\bar{B}_{\bar{3}}}m_{B_{\bar{3}}}} }\mathbb{V}^\mu\bar{B}_{\bar{3}}\overleftrightarrow{\partial}_\mu B_{\bar{3}},\nonumber\\
{\cal L}_{B_{\bar{3}}B_{\bar{3}}\sigma}&=&i\ell_B \sigma \bar{B}_{\bar{3}}B_{\bar{3}},\nonumber\\
{\cal L}_{BB_{\bar{3}}\mathbb{P}}
    &=&-i\frac{g_4}{f_\pi} \sum_i\bar{B}_i^\mu \partial_\mu \mathbb{P} B_{\bar{3}}+{\rm H.c.},\nonumber\\
{\cal L}_{BB_{\bar{3}}\mathbb{V}}    &=&\sqrt{2\over m_{\bar{B}}m_{B_{\bar{3}}}}{g_\mathbb{V}\lambda_I} \epsilon^{\mu\nu\lambda\kappa} \partial_\lambda \mathbb{V}_\kappa\sum_i\bar{B}_{i\nu} \overleftrightarrow{\partial}_\mu
   B_{\bar{3}}+{\rm H.c.}.\label{LB}
\end{eqnarray}
 The coupling constants involve in the above Lagrangians should be determined to constrain the Lagrangians.   In Table~\ref{coupling}, we list the  values of these coupling constants used in the calculation, which are cited from the literature~\cite{Chen:2019asm,Liu:2011xc,Isola:2003fh,Falk:1992cx}.
\renewcommand\tabcolsep{0.165cm}
\renewcommand{\arraystretch}{1.5}
\begin{table}[h!]
\caption{The parameters and coupling constants adopted in our
calculation. The $\lambda$ and $\lambda_{S,I}$ are in the unit of GeV$^{-1}$. Others are in the unit of $1$.
\label{coupling}}
\begin{tabular}{cccccccccccccccccc}\bottomrule[2pt]
$\beta$&$g$&$g_V$&$\lambda$ &$g_{s}$\\
0.9&0.59&5.9&0.56 &0.76\\\hline
$\beta_S$&$\ell_S$&$g_1$&$\lambda_S$ &$\beta_B$&$\ell_B$ &$g_4$&$\lambda_I$\\
-1.74&6.2&-0.94&-3.31&$-\beta_S/2$&$-\ell_S/2$&$g_1/{2\sqrt{2}\over 3}$&$-\lambda_S/\sqrt{8}$ \\
\bottomrule[2pt]
\end{tabular}
\end{table}

In the calculation,  the masses of  particles  are chosen as suggested values in the Review of  Particle Physics  (PDG)~\cite{Tanabashi:2018oca}. The mass difference from the charge is neglected, and average mass is adopted. For example,  the mass of the $\bar{D}$ meson is chosen as $(m_{\bar{D}^0}+m_{D^-})/2$.  The effect of such treatment is negligible on the result and conclusion of this work.  For the broad $\sigma/f_0(500)$ meson, only a range of the pole, $(400-550)-i(200-350)$,  is provided in  PDG. Here, we adopt a mass of 500 MeV. The different choices of the mass of $\sigma$ meson from 400 to 550 MeV will effect the result a little, and can be smeared by a small variation of the cutoff. 

\subsection{Potential of $\Sigma^{(*)}_c\bar{D}^{(*)}-\Lambda_c\bar{D}^{(*)}$   interaction}

The potential of the $\Sigma^{(*)}_c\bar{D}^{(*)}-\Lambda_c\bar{D}^{(*)}$   interaction can be constructed with the help of the vertices for the heavy meson/baryon and the exchanged light meson, which can be easily obtained from the  above Lagrangians.  Besides the vertices,   the propagators of the exchanged light mesons are also needed, which read,
\begin{eqnarray}%
P_\mathbb{P}(q^2)&=& \frac{i}{q^2-m_\mathbb{P}^2}~f_i(q^2),\nonumber\\
P^{\mu\nu}_\mathbb{V}(q^2)&=&i\frac{-g^{\mu\nu}+q^\mu q^\nu/m^2_{\mathbb{V}}}{q^2-m_\mathbb{V}^2}~f_i(q^2),\nonumber\\
P_\sigma(q^2)&=&\frac{i}{q^2-m^2_\sigma}~f_i(q^2),
\end{eqnarray}
where the form factor $f_i(q^2)$ is adopted to compensate the
off-shell effect of exchanged meson. In this work, we  introduce four types of from factors to check the effect of the form factor on the results, which are in forms of  
\begin{align}
f_1(q^2)&=\frac{\Lambda_e^2-m_e^2}{\Lambda_e^2-q^2},\label{FF1}\\
f_2(q^2)&=\frac{\Lambda_e^4}{(m_e^2-q^2)^2+\Lambda^4},\\
f_3(q^2)&=e^{-(m_e^2-q^2)^2/\Lambda_e^2},\\
f_4(q^2)&=\frac{\Lambda_e^4+(q^2_t-m_e^2)^2/4}{[q^2-(q^2_t+m_e^2)/2]^2+\Lambda_e^4},\label{FF4}
\end{align} 
where $m_e$ and $q$ are the mass and momentum of the exchanged light meson. The  $q_t^2$ denotes the value of $q^2$ at the kinematical threshold. The cutoff is rewritten as a form of $\Lambda_e=m+\alpha_e~0.22$ GeV.  In the calculation we also consider the propagators without a form factor, we remark it as $f_0(q^2)=1$.

Because six channels are considered in the current work, it is 
tedious and fallible  to give the explicit of  36 potential elements for the potential of the coupled-channel interaction and input them into the code.  Instead, in this work, we input  the vertices $\Gamma$ and the above propagators $P$  into the code directly and the potential can be obtained as
\begin{eqnarray}%
{\cal V}_{\mathbb{P},\sigma}=f_I\Gamma_1\Gamma_2 P_{\mathbb{P},\sigma}(q^2),\quad
{\cal V}_{\mathbb{V}}=f_I\Gamma_{1\mu}\Gamma_{2\nu}  P^{\mu\nu}_{\mathbb{V}}(q^2).
\end{eqnarray}
Hence,  the explicit forms of the potentials are not given here.
The $f_I$ is the flavor factor for certain meson exchange of certain interaction. It can be derived with the Lagrangians in Eqs.~(\ref{LD}) and (\ref{LB}) and the matrices in Eqs.~(\ref{MPV}) and (\ref{MBB}). The explicit values are listed in Table~\ref{flavor factor}.
\renewcommand\tabcolsep{0.27cm}
\renewcommand{\arraystretch}{1.5}
\begin{table}[h!]
\caption{The flavor factors for certain meson exchanges of certain interaction. The values in bracket are for the case of $I=3/2$ if the values are different from these of $I=1/2$.  \label{flavor factor}}
{	\begin{tabular}{c|ccccc}\bottomrule[2pt]
& $\pi$&$\eta$&$\rho$ & $\omega$ & $\sigma$ \\\hline
$\bar{D}^{(*)}\Sigma^{(*)}_c\to\bar{D}^{(*)}\Sigma^{(*)}_c$&$-1[\frac{1}{2}]$ &$\frac{1}{6}[\frac{1}{6}]$ &$-1[\frac{1}{2}]$&$\frac{1}{2}[\frac{1}{2}]$ & 1\\
$\bar{D}^{(*)}\Lambda_c\to\bar{D}^{(*)}\Lambda_c$&$0$ &$0$ &$0$&$1$ & 2\\
$\bar{D}^{(*)}\Lambda_c\to\bar{D}^{(*)}\Sigma_c^{(*)}$&$\sqrt{6}\over{2}$ &$0$ &$\sqrt{6}\over{2}$&$0$ & 0\\
\toprule[2pt]
\end{tabular}}

\end{table}

\subsection{The qBSE approach}
The scattering amplitude can be calculated with the help of the potential of the interaction obtained in the above.  The Bethe-Salpeter equation is widely used to treat two-body scattering.   With a quasipotential approximation, the 4-dimensional Bethe-Salpeter equation can be reduced to a 3-dimensional  equation and the unitary is kept. As our previous works~\cite{He:2015cea,He:2014nya,He:2015mja,He:2012zd,He:2015yva,He:2017aps}, a spectator approximation, which was explained explicitly in the appendices of Ref.~\cite{He:2015mja}, will be adopted in this work to search the possible bound states.  A bound state from the interaction corresponds to a pole of the scattering amplitude ${\cal M}$.

After partial-wave decomposition,  the 3-dimensional Bethe-Saltpeter equation  after spectator quasipotential approximation can be reduced further to a 1-dimensional  equation with fixed spin-parity $J^P$ as~\cite{He:2015mja},
\begin{eqnarray}
i{\cal M}^{J^P}_{\lambda'\lambda}({\rm p}',{\rm p})
&=&i{\cal V}^{J^P}_{\lambda',\lambda}({\rm p}',{\rm
p})+\sum_{\lambda''}\int\frac{{\rm
p}''^2d{\rm p}''}{(2\pi)^3}\nonumber\\
&\cdot&
i{\cal V}^{J^P}_{\lambda'\lambda''}({\rm p}',{\rm p}'')
G_0({\rm p}'')i{\cal M}^{J^P}_{\lambda''\lambda}({\rm p}'',{\rm
p}),\quad\quad \label{Eq: BS_PWA}
\end{eqnarray}
where the sum extends only over nonnegative helicity $\lambda''$. 
Here, the
reduced propagator with the spectator approximation can be written down in the center-of-mass frame with $P=(W,{\bm 0})$ as
\begin{align}
	G_0&=\frac{\delta^+(p''^{~2}_h-m_h^{2})}{p''^{~2}_l-m_l^{2}}
	\nonumber\\&=\frac{\delta^+(p''^{0}_h-E_h({\rm p}''))}{2E_h({\rm p''})[(W-E_h({\rm
p}''))^2-E_l^{2}({\rm p}'')]}.
\end{align}
Here, as required by the spectator approximation, the heavier particle (remarked with $h$)  is on shell, which satisfies  $p''^0_h=E_{h}({\rm p}'')=\sqrt{
m_{h}^{~2}+\rm p''^2}$. The $p''^0_l$ for the lighter particle (remarked as $l$) is then $W-E_{h}({\rm p}'')$. Here and hereafter, a definition ${\rm p}=|{\bm p}|$ will be adopted.

The partial wave potential is defined with the potential of the interaction obtained in the above as
\begin{eqnarray}
{\cal V}_{\lambda'\lambda}^{J^P}({\rm p}',{\rm p})
&=&2\pi\int d\cos\theta
~[d^{J}_{\lambda\lambda'}(\theta)
{\cal V}_{\lambda'\lambda}({\bm p}',{\bm p})\nonumber\\
&+&\eta d^{J}_{-\lambda\lambda'}(\theta)
{\cal V}_{\lambda'-\lambda}({\bm p}',{\bm p})],
\end{eqnarray}
where $\eta=PP_1P_2(-1)^{J-J_1-J_2}$ with $P$ and $J$ being parity and spin for system, $\bar{D}^{(*)}$ meson or $\Sigma_c^{(*)}$ baryon. The initial and final relative momenta are chosen as ${\bm p}=(0,0,{\rm p})$  and ${\bm p}'=({\rm p}'\sin\theta,0,{\rm p}'\cos\theta)$. The $d^J_{\lambda\lambda'}(\theta)$ is the Wigner d-matrix.

 One may note that we make the partial wave decomposition on the spin parity $J^P$, and the explicit orbital angular momentum $L$ is not involved here.  It is consistent with relativistic treatment adopted in the qBSE approach because the $L$ is not a good quantum number in a relativistic theoretical frame.  With such treatment,  the contributions form all partial waves based on orbital angular momentum $L$ related to a certain $J^P$ considered have been  included already. It is an advantage of our method because the experiment result is usually provided with spin parity $J^P$.  Hence, in this work,  the S-wave state means that a state  can couple to two constituent particles,  the $\Sigma^{(*)}_c\bar{D}^{(*)}-\Lambda_c\bar{D}^{(*)}$ here,  in S wave while all other possible higher partial waves on $L$ are included naturally.

Now we need treat an integral equation,  to avoid divergence,  a regularization is usually introduced. For example, a cutoff in momentum is introduced as one way to do the regularization  in the chiral unitary approach~\cite{Oset:1997it}, which is
related to the dimensional regularization~\cite{Oller:1998hw}.  In the qBSE approach,  we usually adopt an  exponential
regularization  by introducing a form factor into the propagator as
\begin{eqnarray}
G_0({\rm p})\to G_0({\rm p})\left[e^{-(k_l^2-m_l^2)^2/\Lambda_r^4}\right]^2,\label{regularization}
\end{eqnarray} 
where $k_l$ and $m_l$ are the momentum and mass of  the lighter one of meson and baryon.  The interested reader is referred to Ref.~\cite{He:2015mja} for further information about the regularization.  In Ref.~\cite{Liu:2019zvb},  the authors warned that the $\pi$ exchange provides excessive short-range interaction. In the current work,  the relation of the cutoff $\Lambda_r=m+\alpha_r~0.22$ GeV with $m$ being the mass of the exchanged meson is also introduced into the regularization form factor as in those for the  exchanged mesons. Such treatment will suppress the large-momentum, $i.\ e.$, the short-range contribution of the $\pi$ exchange. 

The 1-dimensional integral equation can be easily transformed into a matrix equation. The pole of scattering  amplitude $\cal M$ can be searched by variation of $z$ to satisfy
$|1-V(z)G(z)|=0$
with  $z=W+i\Gamma/2$ equaling to the system energy $W$ at the
real axis~\cite{He:2015mja}.

\section{Single-channel results}
The coupled-channel effect should be included into physical scattering. However, for the $\Sigma^{(*)}_c\bar{D}^{(*)}-\Lambda_c\bar{D}^{(*)}$  interaction considered in the current work,  the coupled-channel effect should be small because the experimental pentaquarks are close to the thresholds. Our previous work~\cite{He:2019ify} also supports such judgement.  Moreover, the coupled-channel effect will make the results complex, which makes it difficult to show the property of bound states from each channels.  Here, we will first present the results of a single-channel calculation.

\subsection{Isodoublet bound states with $I=1/2$}
\begin{figure}[h!]
\centering
\includegraphics[scale=1.7]{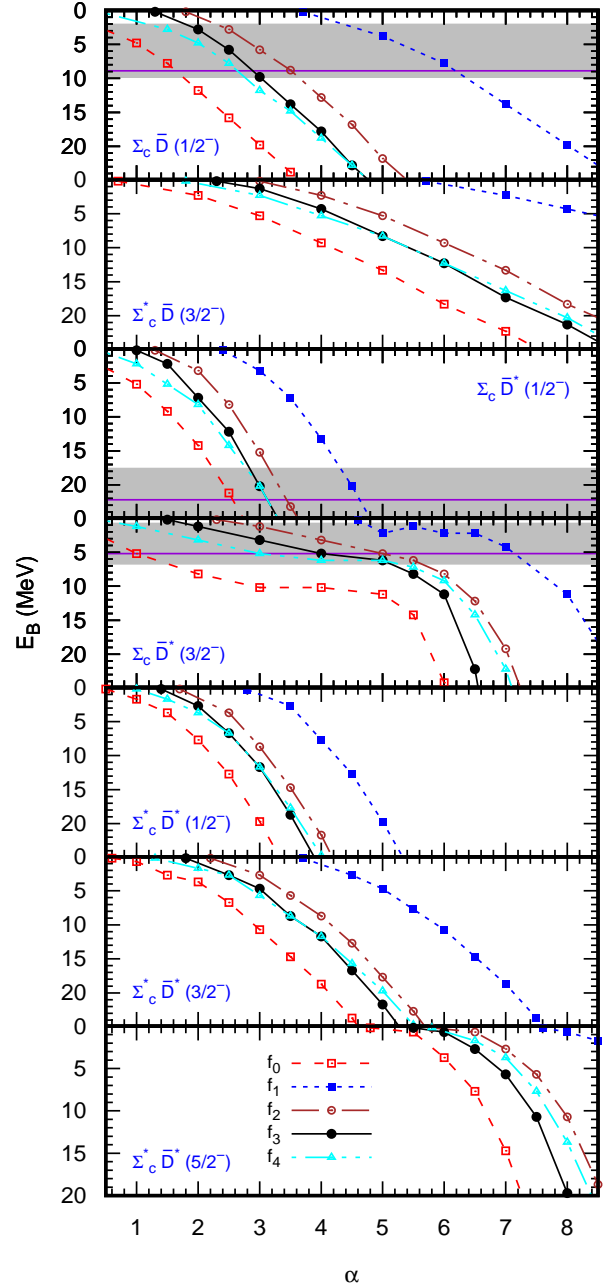}
\caption{The binding energy $E_B$ with the variation of the $\alpha$ for isodoublet bound states from the single-channel interaction.  The thresholds $M_{th}$ for the $\Sigma_c\bar{D}$, $\Sigma_c^*\bar{D}$ and $\Sigma^*_c\bar{D}^*$ channels are 4320.8, 4385.3, 4462.2 and 4526.7 MeV, respectively. The $f_i$ with $i=0,\ 1, \ 2,\ 3,\ 4$ means the results without form factor for exchanged meson or with form factor $f_i(q^2)$ in Eqs.~(\ref{FF1}-\ref{FF4}), respectively. The horizontal lines and the bands are the experimental mass and uncertainties observed at LHCb~\cite{Aaij:2019vzc}.\label{massI1}}
\end{figure}

Now, we consider the isodoublet bound states from single-channel interaction.  In the current work, we have two free parameters,  cutoff parameters $\alpha_r$ and $\alpha_e$ for the regularization and the exchanged meson, respectively.  In single-channel calculation here, we take $\alpha_r=\alpha_e=\alpha$ for simplification. Since the cutoff $\Lambda$ should be about 1 GeV,  we vary $\alpha$ in a range from 0.5 to 8.5 to find the bound state from each channel, which exhibits as a pole in the real axis of the complex plane of $z$. The obtained binding energies $E_B$with the variation of the $\alpha$ are illustrated in Fig.~\ref{massI1}. Here, binding energy is defined as $E_B=M_{th}-W$ with $M_{th}$ and $W$ being the threshold and $W$ of the pole.

In the current calculation, we consider ten  interactions, $\Sigma_c\bar{D}$ with spin parity $1/2^-$, $\Sigma_c^*\bar{D}$ with $3/2^-$, $\Sigma_c\bar{D}^*$ with $1/2^-$ and $3/2^-$,  $\Sigma^*_c\bar{D}^*$ with $1/2^-$,  $3/2^-$, and $5/2^-$, $\Lambda_c \bar{D}^*$ with $1/2^-$ and $3/2^-$,  and $\Lambda_c\bar{D}$ with $1/2^-$.  With  reasonable parameters,  no bound state can be  produced from the $\Lambda_c\bar{D}^{*}$  and $\Lambda_c\bar{D}$ interactions. For other seven interactions, the bound states are produced in the range of the $\alpha$ considered in the current calculation. Here, the results with different types of form factors for exchanged meson  are presented in Fig.~\ref{massI1}. The results show that the different choices of the form factors change the quantitive results but the qualitative conclusion are not changed.  Moreover, the order of the curves for different form factors are almost the same for the seven bound states. It indicates that if appropriate cutoffs are adopted, the same conclusion can be reached with different choices of the form factors.

The $\Sigma_c\bar{D}$ bound state  and two $\Sigma_c\bar{D}^*$ bound states can be related to the LHCb pentaquarks.  Because there is only one S-wave bound state, the $P_c(4312)$ should be assigned into the $\Sigma_c\bar{D}(1/2^-)$ state in the molecular state picture.  Two pentaquarks, $P_c(4457)$ and $P_c(4440)$ were observed at LHCb near the $\Sigma_c\bar{D}^*$ 
threshold, which can be related to two bound states with $1/2^-$ and $3/2^-$ from the $\Sigma_c\bar{D}^*$ interaction.  For the $\Sigma_c\bar{D}^*(3/2^-)$ state, the binding energy increases to about 20 MeV at $\alpha=2-3$ for form factor  $f_{(0, 2,3,4)}$ and $\alpha$ of about 5 for $f_1$.  For the $\Sigma_c\bar{D}^*(1/2^-)$ state,  in a large range of $\alpha$, from 1 to about 6, the binding energy is smaller than 10 MeV. Such results suggest we should assign the $1/2^-$ state as $P_c(4440)$ and the $3/2^-$ state as $P_c(4457)$ state.  Compared with experimental results, the $\alpha$ should be about 3 for $f_{(2,3,4)}$ and about 5  for $f_1$.
With such choice, the $\Sigma_c\bar{D}(1/2^-)$ state has a binding energy about 10 MeV, which is quite consistent with the experimental value.  The results also suggest that the form factor $f_{(2,3,4)}$ is more suitable to explain the  three LHCb pentaquarks in the molecular state picture in the single-channel calculation. 

Based on the above analysis, though the other four bound states, $\Sigma_c^*\bar{D}(3/2^-)$, $\Sigma^*_c\bar{D}^*(1/2^-)$,  $\Sigma^*_c\bar{D}^*(3/2^-)$,  $\Sigma^*_c\bar{D}^*(5/2^-)$, can be produced with variation of the cutoff, the existence of the $\Sigma^*_c\bar{D}^*(5/2^-)$ should be doubted because an $\alpha$ larger than 5 is required to produce such state, which is much larger the one to produce three LHCb pentaquarks with the experimental masses at the same time.   If we adopt $\alpha=3$  for the form factor $f_{(2,3,4)}$, the binding energies of $\Sigma_c^*\bar{D}^*(1/2^-)$ and $\Sigma_c^*\bar{D}^*(3/2^-)$  states are about 5 MeV, and the $\Sigma_c^*\bar{D} (3/2^-)$ state should have a very small binding energy.

\subsection{Isoquartet bound states with $I=3/2$}

\begin{figure}[htpb!]
\centering
\includegraphics[scale=1.7]{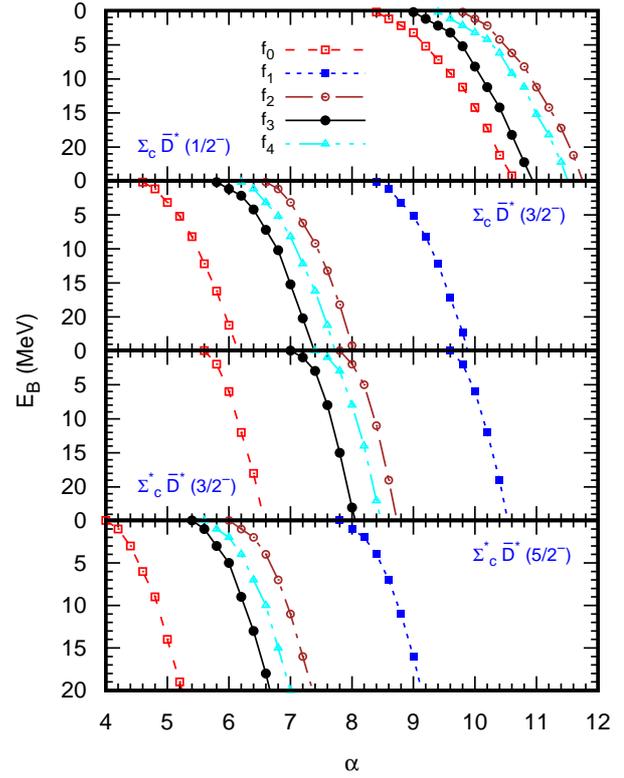}
\caption{The binding energy $E_B$ with the variation of the $\alpha$ for isoquartet bound states. The other
conventions are the same as in Fig. \ref{massI1}.}
\label{massI3}
\end{figure}

In the above, we present the isodoublet bound states.  For the same interaction with different isospins, the model and parameters involved should be also the same. Hence,  it is straight forward to give the prediction about the isoquartet bound states from the $\Sigma^{(*)}_c\bar{D}^{(*)}$ and $\Lambda_c\bar{D}^{(*)}$  interactions.  The possible experimental observation about such bound states is also a good check to the molecular state interpretation of the LHCb pentaquarks and the results in this work.  Here, we make the calculation to search  isoquartet bound states in the same model as in the isodoulet case.  The results  are presented in Fig~\ref{massI3}.

Here, we still vary the $\alpha$ to search for  the bound states from the interactions as in the isodoublet  case.  
Generally speaking, a larger $\alpha$ should be adopted to produce the bound states. If we focus on the results with $f_{(2,3,4)}$, which is more suitable to reproduce the LHCb pentaquarks, the $\alpha$ is at least larger than 5 to produce a bound state from the interaction considered here.  With an $\alpha$ smaller than 12, no bound state can be found for the $\Sigma_c\bar{D}(1/2^-)$,  the $\Sigma^*_c\bar{D}(3/2^-)$, and the $\Sigma^*_c\bar{D}^*(1/2^-)$ interactions.  For the $\Sigma_c \bar{D}^*(1/2^-)$ interaction, an $\alpha$ larger than 8 is required to force the potential strong enough to produce a bound state.  For the $\Sigma_c\bar{D}^*(3/2^-)$ and $\Sigma^*_c\bar{D}^*(3/2^-)$ interactions,  the bound states are produced at an $\alpha$ of about 6 GeV.  The bound state from $\Sigma^*\bar{D}^*(5/2^-)$ appears at an $\alpha$ of about 5 for $f_{(2,3,4)}$ and more larger for $f_1$.
If we recall that  the LHCb pentaquarks are reproduced at an $\alpha$ of 3 for $f_{(2,3,4)}$ and 5 for $f_1$ in the isodoublet case, it is reasonable to doubt the existence of the four bound states shown in Fig~\ref{massI3} if the assignment of three LHCb pentaquarks are right.

Usually,  increase of the $\alpha$ can enhance the strength of the interaction. The large $\alpha$ required here suggest that the isoquaret interactions are much weaker than those with $I=1/2$.  It is easy to understand if we recall the flavor factors listed in Table 2. The sign between the potentials by  $\pi$ and $\eta$ exchanges, and that between $\rho$ and $\omega$ exchanges is different for $I=1/2$ and $3/2$ cases. It results in the cancellation of two contributions in the isoquaret case.   Such cancellation makes the interaction with $I=3/2$ too weak to produce a bound state with a small $\alpha$.

\section{Coupled-channel results}

It is well known that  the coupled-channel effect will affect the binding energy of the bound state. Moreover, if a lower channel was considered, the bound state from the channel with higher threshold will acquire a width. In the above single-channel calculation, six bound states are produced [the $\Sigma_c^*\bar{D}(5/2^-)$ is not well supported because a large $\alpha$ is required].  Those states are from three channels, which can be coupled to each other by   light meson exchange as the single-channel interaction.  With the Lagrangians in the heavy quark limit and chiral symmetry, a coupled-channel calculation can be preformed  for the $\Sigma^{(*)}_c\bar{D}^{(*)}-\Lambda_c\bar{D}^{(*)}$ interaction. 

\subsection{The poles from the $\Sigma^{(*)}_c\bar{D}^{(*)}-\Lambda_c\bar{D}^{(*)}$ interaction}

\begin{figure}[htpb!]
\centering
\includegraphics[scale=1.0]{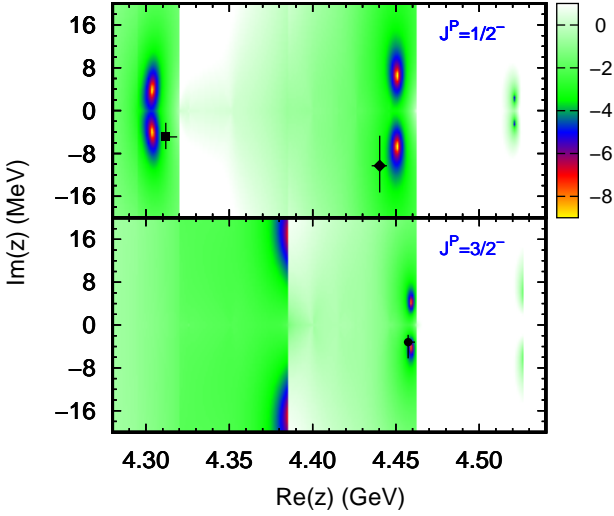}
\caption{The $\log|1-V(z)G(z)|$ with the variation of  $z$ for the $\Sigma^{(*)}_c\bar{D}^{(*)}-\Lambda_c\bar{D}^{(*)}$ interaction with $J^P=1/2^-$ and $3/2^-$ at $\alpha=2.5$. The  color  means the value of $\log|1-V(z)G(z)|$ as shown in the color box. The form factor $f_3$ is adopted in the calculation. The full square, diamond, and  circle are for the experimental data of the $P_c(4312)$, $P_c(4440)$ and $P_c(4457)$ at LHCb~\cite{Aaij:2019vzc}}
\label{cc1}
\end{figure}

In single-channel calculation, the bound state is a pole at real axis. After the coupled-channel effect are included, the pole will leave the real axis and becomes a pole in the complex plane as shown in Fig.~\ref{cc1}. The poles from the $\Sigma^{(*)}_c\bar{D}^{(*)}-\Lambda_c\bar{D}^{(*)}$ interaction with $J^P=1/2^-$ and $3/2^-$ are presented in the figure. 
 
Here we take the $f_3$ to show the coupled channel results. The results with $f_{(1,2,4)}$ is qualitatively consistent with the results with $f_3$ if the cutoff is varies correspondingly. 
 In the single-channel calculation, the best value of  $\alpha$ is found  about 3 to reproduce the experimental masses of three LHCb pentaquarks.  After including the coupled-channel effect, besides the bound states have width and become  resonances, the masses of the resonances are also different from the masses obtained from the single-channel calculation. If we still adopt an $\alpha$ of 3, the pole of  the $\Sigma_c\bar{D}(1/2^-)$ state will move from  $4311$ to $4294$~MeV, which is even below the $\Lambda_c\bar{D}^*$ threshold.  Hence, we adopt a smaller value $\alpha$ of $2.5$ to give the poles from  the $\Sigma^{(*)}_c\bar{D}^{(*)}-\Lambda_c\bar{D}^{(*)}$ interaction.

Six poles can be found in the complex plane with their conjugate partners.  In the case with $J^P=1/2^-$, there exist three poles near the $\Sigma_c\bar{D}$, $\Sigma_c\bar{D}^*$ and $\Sigma^*\bar{D}^*$ thresholds, respectively.  In the case with $J^P=3/2^-$, we also have three poles near the $\Sigma^*_c\bar{D}$, $\Sigma_c\bar{D}^*$ and $\Sigma^*\bar{D}^*$ thresholds, respectively.
The pole near the $\Sigma_c\bar{D}$ threshold is obviously related to the $P_c(4312)$. Compared with  experimental values at LHCb,  $M=4311.9\pm 0.7_{-0.6}^{+6.8}$ and $\Gamma=9.8\pm 2.7_{-4.5}^{+3.7}$ MeV, the theoretical pole at $4304-4i$ MeV is a little lower but in the uncertainties of  the width  [here we use the relation $\Gamma=-2$ Im$(z)$ ].   Two poles appear near the $\Sigma_c\bar{D}^*$ with $1/2^-$ and $3/2^-$, respectively.  The pole in $3/2^-$ fall in the experimental values with uncertainties quite well while the $1/2^-$ pole is a little higher in mass. Though the mass gap of these two states is narrowed after the coupled-channel effect included, the order of these two states still supports the assignment of two states with $1/2^-$ and $3/2^-$ as $P_c(4440)$ and $P_c(4457)$, respectively. The theoretical widths also support such assignment. Hence, as the single-channel results, the coupled-channel results support the assignment of the $P_c(4457)$, $P_c(4440)$ and $P_c(4312)$ as molecular states $\Sigma_c\bar{D}(1/2^-)$, $\Sigma_c\bar{D}^*(1/2^-)$ and $\Sigma_c\bar{D}(3/2^-)$, respectively. 

It is interesting to observe a pole near the $\Sigma^*\bar{D}$ threshold, which may be related to the $P_c(4380)$ suggested in the old LHCb article~\cite{Aaij:2015tga}.  This pole is almost on the threshold, if the physical strength of the interaction is a little weaker, it may become a cusp on the threshold.  Besides, it has a width about 40 MeV, which is much larger than  three LHCb pentaquarks.  These properties of this state may be the reason why the $P_c(4380)$ is very broad and difficult to be observed in the invariant mass spectrum.  

Near the $\Sigma^*\bar{D}^*$ threshold,  two poles can be found with both $1/2^-$ and $3/2^-$.  The $1/2^-$ pole is at $4521-2i$ and $3/2^-$ pole is at $4526-6i$.  These two poles are obviously shallower than other poles. It indicates that the peaks corresponding of these states may be smaller than other states.

\subsection{The widths of the molecular states}

From the above results, we can find that the widths of the $P_c$ states can be well reproduced in our model. It is interesting to give the widths from each channel to show the strength of the coupling between the molecular state and the corresponding channel.
In the current work, the pole of a molecular in the complex plane can be obtained by a coupled-channel calculation. Usually, the width of a state can be obtained as $\Gamma=-2{\rm Im}(z)$ where the $z$ is the position of the pole, that is, the width can be related to the imaginary part of the pole.  In Table~\ref{pole},  we list the poles of the molecular states for full coupled-channel calculation and two-channel calculation.

\renewcommand\tabcolsep{0.2cm}
\renewcommand{\arraystretch}{1.31}
\begin{table*}[hpbt!]
\caption{The positions and branching ratios  of the molecular states. The ``$CC$" means the full coupled-channel calculaiton.  The ``pole"  means mass of corresponding threshold subtracted by the position of a pole, $M_{th}-z$,  in the unit of MeV and Br$=\frac{{\rm Im}_{i}}{\sum {\rm Im}_i}$  for $i$ channel in the  unit of $\%$. The $\alpha_r$ is the cutoff in the exponential regularization in Eq.~(\ref{regularization}).  The explicit explanation can be found in the text.
\label{pole}}
\begin{tabular}{r|r|rrrrrrrrrr|rr}\hline 
$\alpha_r$ &\multicolumn{1}{c|}{$CC$}& \multicolumn{2}{c}{$\Sigma_c \bar{D}^*$} & \multicolumn{2}{c}{$\Sigma^*_c \bar{D}$} &  \multicolumn{2}{c}{ $\Sigma_c \bar{D}$}&   \multicolumn{2}{c}{$\Lambda_c \bar{D}^*$} &  \multicolumn{2}{c}{  $\Lambda_c \bar{D}$} &\multicolumn{2}{c}{ sum}\\
\hline
    &\multicolumn{1}{c|}{pole}& \multicolumn{1}{c}{pole} &\multicolumn{1}{c}{Br}& \multicolumn{1}{c}{pole} & \multicolumn{1}{c}{Br}& \multicolumn{1}{c}{pole} & \multicolumn{1}{c}{Br}& \multicolumn{1}{c}{pole} & \multicolumn{1}{c}{Br}& \multicolumn{1}{c}{pole} & \multicolumn{1}{c|}{Br}& \multicolumn{1}{c}{$\sum$Im$_i$ }& \multicolumn{1}{c}{$\frac{{\rm Im}_{CC}}{\sum {\rm Im}_i}$ } \\
\hline
\multicolumn{12}{c}{$\Sigma^*_c\bar{D}^*(1/2^-)$}\\
\hline
$1.5$ &$1.2+1.0i$   &$1.8+0.1i$ &$17$    &$2.1+0.1i$  &$17$   &$1.2+0.3i$ &$50$ &$1.7+0.1i$ &$17$ &$1.9+0.0i$  &$ 0$ & $0.6$ & 167\\
$2.0$ &$3.0+1.6i$   &$3.7+0.2i$ &$18$    &$3.9+0.2i$  &$18$   &$2.9+0.5i$ &$45$ &$4.6+0.2i$ &$18$ &$3.7+0.0i$  &$0$ &$1.1$ &145\\
$2.5$ &$5.5+2.3i$   &$6.1+0.3i$ &$19$    &$6.6+0.3i$  &$19$   &$5.3+0.7i$ &$44 $ &$6.4+0.3i$ &$19$ &$6.4+0.0i$  &$0$ & $1.6$ &144\\
$3.0$ &$7.4+3.1i$   &$8.8+0.4i$ &$ 18$    &$9.0+0.4i$  &$18$   &$7.1+1.0i$ &$45$ &$8.4+0.4i$ &$ 10$ &$9.2+0.0i$  &$0 $ & $2.2$ &141 \\
\hline
\multicolumn{12}{c}{$\Sigma^*_c\bar{D}^*(3/2^-)$}\\
\hline
$2.0$ &$0.0+4.2i$  &$0.3+0.7i$  &$28$  &$0.5+0.7i$   &$28$   &$1.2+0.0i$ &$0$ &$0.0+0.9i$ &$36$ &$1.1+0.2i$ &$8$&$2.5$&$168$ \\
$2.5$ &$0.0+5.8i$  &$1.0+1.2i$  &$32$  &$1.1+0.8i$   &$22$   &$2.3+0.0i$ &$0$ &$0.0+1.5i$ &$41$ &$2.3+0.2i$ &$$5&$3.7$&$158$\\
$3.0$ &$0.0+6.8i$  &$1.7+1.7i$  &$37$  &$1.6+1.0i$   &$22$   &$3.4+0.0i$ &$0$ &$0.0+1.7i$ &$37$ &$3.4+0.2i$ &$4$&$4.6$&$148$ \\
$3.5$ &$0.0+7.5i$  &$2.2+2.1i$  &$41$  &$2.0+1.1i$   &$22$   &$4.2+0.1i$ &$2 $ &$0.0+1.5i$ &$29$ &$4.4+0.3i$ &$6$&$5.1$&$147 $ \\
\hline
\multicolumn{12}{c}{$\Sigma_c\bar{D}^*(1/2^-)$}\\
\hline
$1.0$ &$3.5+1.9i$    &\multicolumn{1}{c}{$--$ }     &\multicolumn{1}{c}{$--$ }  &$3.0+0.0i $   &$ 0 $  &$2.9+0.3i  $ & $20 $   &$3.3+1.2i  $   & $80 $ &$3.0+0.0i  $  &$0$  &$1.5 $ & $127$ \\
$2.0$ &$8.2+4.8i$    &\multicolumn{1}{c}{$--$ }     &\multicolumn{1}{c}{$--$ }  &$8.7+0.2i $   &$ 5$  &$8.0+0.5i  $ & $12$   &$9.1+3.3i  $   & $80 $ &$8.7+0.1i  $  &$2$  &$4.1$ & $117$ \\
$3.0$ &$13.8+8.8i$   &\multicolumn{1}{c}{$--$ }     &\multicolumn{1}{c}{$--$ }  &$15.2+0.9i$   &$11 $  &$14.1+0.8i $ & $9 $   &$15.5+6.3i $   & $ 74 $ &$16.2+0.5i $  &$6$  &$8.5 $ & $104$ \\
$4.0$ &$17.7+14.7i$  &\multicolumn{1}{c}{$--$ }     &\multicolumn{1}{c}{$--$ }  &$23.2+2.1i$   &$15 $  &$19.0+1.5i $ & $11 $   &$21.5+9.4i $   & $66  $ &$22.1+1.2i $  &$9$  &$14.2$ & $104$ \\
\hline
\multicolumn{12}{c}{$\Sigma_c\bar{D}^*(3/2^-)$}\\
\hline
$1.0$ &$2.7+1.0i$ &\multicolumn{1}{c}{$--$ }     &\multicolumn{1}{c}{$--$ }  &$1.8+0.3i$ &$19$  &$1.6+0.0i$ &$0$   &$1.4+1.0i$ &$63$ &$1.6+0.3i$ &$19$ &$1.6 $&$63$\\
$1.5$ &$3.5+2.3i$ &\multicolumn{1}{c}{$--$ }     &\multicolumn{1}{c}{$--$ }  &$2.1+0.4i$ &$13$  &$2.0+0.0i$ &$0$   &$0.9+2.4i$ &$75$ &$1.7+0.4i$ &$13$ &$ 3.2$&$ 72$\\
$2.0$ &$3.4+3.6i$ &\multicolumn{1}{c}{$--$ }     &\multicolumn{1}{c}{$--$ }  &$2.1+0.4i$ &$9$  &$2.1+0.1i$ &$2$   &$0.0+3.5i$ &$78 $ &$1.6+0.5i$ &$11$ &$ 4.5$&$80$\\
$2.5$ &$2.8+4.2i$ &\multicolumn{1}{c}{$--$ }     &\multicolumn{1}{c}{$--$ }  &$2.1+0.4i$ &$12$  &$2.0+0.1i$ &$3$   &$0.0+2.4i$ &$71 $ &$1.4+0.5i$ &$15$ &$3.4$ &$81 $ \\
$3.0$ &$2.6+4.5i$ &\multicolumn{1}{c}{$--$ }     &\multicolumn{1}{c}{$--$ }  &$2.0+0.4i$ &$ 13$  &$2.0+0.1i$ &$3$   & $0.0+2.0i$&$65$  &$1.4+0.6i$ & $19$ &$3.1$ &69  \\
\hline 
\multicolumn{12}{c}{$\Sigma_c^*\bar{D}(3/2^-)$}\\
\hline
$2.5$ &$0.0+19i$   &\multicolumn{1}{c}{$--$ }     &\multicolumn{1}{c}{$--$ } &\multicolumn{1}{c}{$--$ }     &\multicolumn{1}{c}{$--$ } &$0.4+0i$  &$0$    &$0.0+16i$ &$100                   $  &$0.4+0i$ &$0$ &$16$ &$ 119$\\
$3.0$ &$0.0+24i$   &\multicolumn{1}{c}{$--$ }     &\multicolumn{1}{c}{$--$ } &\multicolumn{1}{c}{$--$ }     &\multicolumn{1}{c}{$--$ } &$0.6+0i$  &$0$    &$0.0+19i$ &$100                     $  &$0.6+0i$ &$0$ &$19$ &$ 126$ \\
$3.5$ &$0.0+28i$   &\multicolumn{1}{c}{$--$ }     &\multicolumn{1}{c}{$--$ } &\multicolumn{1}{c}{$--$ }     &\multicolumn{1}{c}{$--$ } &$0.9+0i$  &$0$    &$0.0+22i$ &$100                    $  &$0.9+0i$ &$0$ &$22$ &$ 127$ \\
$4.0$ &$0.0+30i$   &\multicolumn{1}{c}{$--$ }     &\multicolumn{1}{c}{$--$ } &\multicolumn{1}{c}{$--$ }     &\multicolumn{1}{c}{$--$ } &$1.0+0i$  &$0$    &$0.0+25i$ &$100$   &$1.0+0i$ &$ 0$ &$25$ &$ 120$ \\
\hline
\multicolumn{12}{c}{$\Sigma_c\bar{D}(1/2^-)$}\\
\hline
$1.0$ &$3.7+2.0i $ &\multicolumn{1}{c}{$--$ }     &\multicolumn{1}{c}{$--$ }&\multicolumn{1}{c}{$--$ }     &\multicolumn{1}{c}{$--$ } &\multicolumn{1}{c}{$--$ }     &\multicolumn{1}{c}{$--$ }  &$3.4+2.1i $&$88$  &$2.1+0.3i$ &$13$ &$2.4$ &$83$\\
$1.5$ &$8.1+2.9i $ &\multicolumn{1}{c}{$--$ }     &\multicolumn{1}{c}{$--$ }&\multicolumn{1}{c}{$--$ }     &\multicolumn{1}{c}{$--$ } &\multicolumn{1}{c}{$--$ }     &\multicolumn{1}{c}{$--$ }  &$6.1+3.0i $&$88$  &$3.3+0.4i$ &$12$ &$3.4$ &$ 85$\\
$2.0$ &$11.4+4.0i$ &\multicolumn{1}{c}{$--$ }     &\multicolumn{1}{c}{$--$ }&\multicolumn{1}{c}{$--$ }     &\multicolumn{1}{c}{$--$ } &\multicolumn{1}{c}{$--$ }     &\multicolumn{1}{c}{$--$ }  &$9.4+4.0i $&$89$  &$4.6+0.5i$ &$11$ &$4.5$ &$89$\\
$2.5$ &$17.8+4.6i$ &\multicolumn{1}{c}{$--$ }     &\multicolumn{1}{c}{$--$ }&\multicolumn{1}{c}{$--$ }     &\multicolumn{1}{c}{$--$ } &\multicolumn{1}{c}{$--$ }     &\multicolumn{1}{c}{$--$ }  &$13.6+4.9i$&$87$  &$5.9+0.7i$ &$13$ &$5.6$ &$82$ \\
$3.0$ &$23.6+4.8i$ &\multicolumn{1}{c}{$--$ }     &\multicolumn{1}{c}{$--$ }&\multicolumn{1}{c}{$--$ }     &\multicolumn{1}{c}{$--$ } &\multicolumn{1}{c}{$--$ }     &\multicolumn{1}{c}{$--$ }  &$18.4+5.1i$&$86$  &$7.1+0.8i$ &$14$ &$5.9$ &$81  $ \\
\hline
\end{tabular}

\end{table*}

From the analysis in section 4.1, one can find that the coupled-channel effects between different channels will effect the position of the poles but not very far. We  can still identify the main contribution of a molecular state from its mass.  Here, we remark the states with their main origin, for example, for the state near $\Sigma^*_c\bar{D}^*$ thresholds with $1/2^-$, we adopt a notation as  $\Sigma^*_c\bar{D}^*(1/2^-)$. In  Table~\ref{pole}, the results for six states $\Sigma^*_c\bar{D}^*(1/2^-)$ $\Sigma_c^*\bar{D}^*(3/2^-)$, $\Sigma_c\bar{D}^*(1/2^-)$, $\Sigma_c\bar{D}^*(3/2^-)$, $\Sigma_c^*\bar{D}(3/2^-)$ and $\Sigma_c\bar{D}(1/2^-)$, which can be produced with an $\alpha$ in a reasonable region where the LHCb pentaquarks can be reproduced, are presented in order.  The positions of the these states with the full coupled-channel calculation are listed in  second column in Table~\ref{pole}. Here, to emphasize the binding energy, we replace the real part of the pole by the binding energy, that is, $z\to M_{th}-z$ with $M_{th}$ being the mass of the threshold. 

In third to twelfth columns, we present the results from the two-channel calculation. In such calculation, we only keep the coupling between  main channel and another channel to study the effect of this channel on the bound state from the  main channel. Again, we take the  $\Sigma^*_c\bar{D}^*(1/2^-)$ given first in  Table~\ref{pole} as example.  For this case, the $\Sigma^*_c\bar{D}^*$ is the main channel. The $\Sigma^*_c\bar{D}^*(1/2^-)$ state is mainly produced from this interaction. If only $\Sigma^*_c\bar{D}^*$ channel is considered, the pole is at the real axis.  After another channel, such as $\Sigma_c\bar{D}^*$,  is added, the pole will move to complex plane. Especially, the imaginary part or width of this state is from the  $\Sigma_c\bar{D}^*$ channel totally in such two-channel calculation.  In table~\ref{pole}, the results for two-channel calculation from  main channel and one of the $\Sigma_c\bar{D}^*$, $\Sigma^*_c \bar{D}$, $\Sigma_c \bar{D}$, $\Lambda_c \bar{D}^*$, and $\Lambda_c \bar{D}$ channels are  given from  third to twelfth columns in order.  Here, we introduce the branching ratio Br to present the importance of corresponding channel. It is defined as the imaginary part of the each channel divided by the sum of the imaginary parts of all channels, that is, ${\rm Br}=\frac{{\rm Im}_{i}}{\sum {\rm Im}_i}$. The results with  different values of  $\alpha_r$ are presented  to show the stability of the branching ratios.  One can find that the sum of the imaginary parts of every channels, listed in the thirteenth column, deviates from the full coupled-channel result, which is shown in the last column as $\frac{{\rm Im}_{CC}}{\sum {\rm Im}_i}$. It is from the couplings between the channels except   the  main channel  $\Sigma^*_c\bar{D}^*$.  Such deviation is small  in all cases. Hence, the branching ratio here should be seen as the first order approximation if the pole of the molecular state is not far away from the threshold of the its main origin.

Two $\Sigma^*_c\bar{D}^*$ states can decay into five channels considered in this work.  In the full coupled-channel calculation, the real and imaginary parts of the pole increase with the increase of $\alpha_r$.  Such behavior can be also found in the two-channel results. However,  the branching ratio of each channel is not sensitive to the variation of the parameter.   The $\Sigma_c\bar{D}$ is the most important decay channel of the $\Sigma^*_c\bar{D}^*(1/2^-)$ state with a branching about 50\%, and the $\Sigma_c\bar{D}^*$, $\Sigma^*_c\bar{D}$ and $\Lambda_c\bar{D}^*$  channels also have considerable contributions with branching ratios a little smaller than 20\%. The $\Lambda_c\bar{D}$ channel has little effect on the decay width of the $\Sigma^*_c\bar{D}^*(1/2^-)$ state (here  and hereafter the $0.0i$ does not means forbidding but   a very small width in the current precision).  For the $\Sigma^*_c\bar{D}^*(3/2^-)$ state, the $\Sigma_c\bar{D}^*$, $\Sigma^*_c\bar{D}$ and $\Lambda_c\bar{D}^*$ channels provide considerable widths with branching ratios about 20-30\%, while its couplings to the $\Sigma_c\bar{D}$ and $\Lambda_c\bar{D}$ channels are very small.

There exist two states near the $\Sigma_c\bar{D}^*$ threshold in our model, which can be related to the experimental $P_c(4440)$ and $P_c(4457)$. The channel above the $\Sigma_c\bar{D}^*$  channel, here $\Sigma^*_c\bar{D}^*$ channel,  does not provide the width, that is, the the pole  near the $\Sigma_c\bar{D}^*$ threshold  from the  two-channel calculation with $\Sigma^*\bar{D}^*$ channel  is still on the real axis.  It reflects that a state $\Sigma\bar{D}^*$  can not decay to $\Sigma^*\bar{D}^*$ which is beyond its mass. Hence, there are only four channels listed.
For both $\Sigma_c\bar{D}^*$ states,   the $\Lambda\bar{D}^*$ channel is dominant, with branching ratio about 70\%. Other channels only have branching ratios smaller than 20\%.  The dominance of the $\Lambda_c\bar{D}^*$ is also found in the $\Sigma_c^*\bar{D}(3/2^-)$ and $\Sigma_c\bar{D}(1/2^-)$ states, where fewer channels are opened in the models considered in the current work.  The branching ratio of the $\Sigma_c^*\bar{D}(3/2^-)$ state to the $\Lambda_c\bar{D}$ channel is 100\% while the $\Lambda_c\bar{D}$ channel provides about 90\% contribution to the $\Sigma_c\bar{D}(1/2^-)$ state.  

\section{Summary and discussion}

In this work, the $\Sigma^{(*)}_c\bar{D}^{(*)}-\Lambda_c\bar{D}^{(*)}$ interaction is studied in the qBSE approach with the help of the Lagrangians in heavy quark limit and with chiral symmetry. The single-channel calculation shows that  three LHCb pentaquarks, $P_c(4312)$, $P_c(4440)$, and $P_c(4457)$  can be well reproduced from the $\Sigma_c\bar{D}$ interaction with spin parity $J^P=1/2^-$ and $\Sigma_c\bar{D}^{*}$ interaction with $1/2^-$ and $3/2^-$, respectively. It is further supported by the coupled-channel calculation, where the bound states become poles in the complex plane, and acquire widths in the uncertainties of the experimental values.  Our results also suggest that the $P_c(4440)$ and $P_c(4457)$ should have large branching ratios in the $\Lambda_c\bar{D}^*$ channel, and this channel is also very important in the decay of the $P_c(4312)$.
The $P_c(4380)$, which is suggested by the first LHCb experiment and suspended in the updated results,  can be related to the $\Sigma_c^*\bar{D}$ state with $3/2^-$. This state is on the  $\Sigma_c^*\bar{D}$ threshold and has a large width from $\Lambda\bar{D}^*$ channel. It may be only a cusp on the threshold. If so, the peak in the invariant mass spectrum of this state will be broad and difficult to be distinguished in experiment. 

Other possible molecular states from the  $\Sigma^{(*)}_c\bar{D}^{(*)}-\Lambda_c\bar{D}^{(*)}$ interaction are also predicted  in the same model.  three $\Sigma_c^*\bar{D}^*$ states can be produced with appropriate  $\alpha$ adopted. However, very large $\alpha$ are required to produce the state with $5/2^-$.  Such results are consistent with the Scenario A of Ref.~\cite{Pan:2019skd}, where the binding energy of the state with $5/2^-$ is the smallest.  If we adopt a value of $\alpha$ which can reproduce three LHCb pentaquark,  only two states, $\Sigma_c^*\bar{D}^*$ with $1/2^-$ and $3/2^-$ are suggested by our results.  The decay patterns of this two states are also studied in the coupled-channel calculation. The $\Sigma_c\bar{D}$ channel is found important to the state with $1/2^-$ and three channels, $\Sigma_c\bar{D}^*$, $\Sigma^*\bar{D}$, and $\Lambda_c\bar{D}$ have considerable contributions to the state with $3/2^-$. The isoquaret molecular states with $I=3/2$ are also studied in the same model.  The results suggest that the interaction is very weak. Only three states, $\Sigma_c\bar{D}(1/2^-)$, $\Sigma^*\bar{D}^*(3/2^-)$, and $\Sigma_c\bar{D}(5/2^-)$ can be produced but very large $\alpha$ are required. 

\vskip 10pt

\noindent {\bf Acknowledgement} This project is supported by the National Natural Science
Foundation of China (Grants No. 11675228, and No. 11375240), and the Fundamental Research Funds for the Central Universities.

%


\end{document}